\documentstyle[prl,aps]{revtex}
\twocolumn
\begin{document}
\author{Jian-Qi Shen $^{1,}$$^{2}$ \footnote{E-mail address: jqshen@coer.zju.edu.cn}}
\address{$^{1}$ Centre for Optical
and Electromagnetic Research, State Key Laboratory of Modern
Optical Instrumentation, \\Zhejiang University,
Hangzhou SpringJade 310027, P.R. China\\
$^{2}$ Zhejiang Institute of Modern Physics and Department of
Physics, Zhejiang University, Hangzhou 310027, P.R. China}
\date{\today }
\title{Comment on ``Solutions of the Schr\"{o}dinger equation for the time-dependent linear potential''}
\maketitle

\begin{abstract}
We show that the solution obtained by Bekkar {\it et al.} in their
comment [Phys. Rev. A {\bf 68}, 016101 (2003)] on Guedes's work of
solving the quantum system with a time-dependent linear potential
is still {\it not} the {\it general} one of the Schr\"{o}dinger
equation. It is concluded that Bekkar {\it et al.}'s solution
(corresponding to the linear Lewis-Riesenfeld invariant) and our
solution (corresponding to the quadratic-form Lewis-Riesenfeld
invariant) presented here will constitute together a complete set
of solutions (general solutions) of the time-dependent
Schr\"{o}dinger equation of the system under consideration.
          \\ \\
PACS number(s): 03.65.Fd, 03.65.Ge
\end{abstract}
\pacs{}

Recently, Guedes used the Lewis-Riesenfeld invariant
formulation\cite{Lewis} and solved the one-dimensional
Schr\"{o}dinger equation with a time-dependent linear
potential\cite{Guedes}, the Hamiltonian of which is
$H(t)=\frac{p^{2}}{2m}+f(t)q$ with $q$ and $p$ being the canonical
variables. More recently, Bekkar {\it et al.} pointed out
that\cite{Bekkar} the result obtained by Guedes is merely the
particular solution (that corresponds to the null eigenvalue of
the linear Lewis-Riesenfeld invariant) rather than a general one.
In the comment\cite{Bekkar}, Bekkar {\it et al.} stated that they
correctly used the invariant method\cite{Lewis} and gave the
general solutions of the time-dependent Schr\"{o}dinger equation
with a time-dependent linear potential\cite{Bekkar}. However, in
the present comment, we will show that although the solutions of
Bekkar {\it et al.} is more general than that of
Guedes\cite{Guedes}, what they finally achieved in their
comment\cite{Bekkar} may be still {\it not} the {\it general}
solutions, either. On the contrary, I think that their result
\cite{Bekkar} might also belongs to the particular one. The reason
for this may be as follows: according to the Lewis-Riesenfeld
invariant method\cite{Lewis}, the solutions of the time-dependent
Schr\"{o}dinger equation can be constructed in terms of the
eigenstates of the Lewis-Riesenfeld (L-R) invariants. It is known
that both the squared of a L-R invariant (denoted by $I(t)$) and
the product of two L-R invariants are also the invariants, which
agree with the Liouville-Von Neumann equation $\frac{\partial
}{\partial t}I(t)+\frac{1}{i}\left[ I(t),H(t)\right] =0$, and that
if $I_{a}$ and $I_{b}$ are the two L-R invariants of a certain
time-dependent quantum system and $|\psi(t)\rangle$ is the
solution of the time-dependent Schr\"{o}dinger equation
(corresponding to one of the invariants, say, $I_{a}$), then
$I_{b}|\psi(t)\rangle$ is another solution of this quantum system.
So, in an attempt to obtain the {\it general} solutions of a
time-dependent system, one should first analyze the complete set
of all L-R invariants of the system under consideration.
Historically, in order to obtain the complete set of invariants,
Gao {\it et al.} suggested the concept of basic invariants which
can generate the complete set of invariants\cite{Gao}, as stated
in Ref.\cite{Gao}, the basic invariants can be called invariant
generators. As far as Bekkar {\it et al.}'s result\cite{Bekkar} is
concerned, the obtained solutions are the ones corresponding only
to the linear invariant ({\it i.e.}, $I_{\rm l
}(t)=A(t)p+B(t)q+C(t)$) that is simply one of the L-R invariants
constituting a complete set. It is apparently seen that the
quadratic form, $I_{\rm
q}(t)=D(t)p^{2}+E(t)(pq+qp)+F(t)q^{2}+A'(t)p+B'(t)q+C'(t)$, is
also the one that can satisfy the Liouville-Von Neumann equation,
since it is readily verified that the generators of $I_{\rm q}(t)$
form a Lie algebra\cite{Shen1}. However, for the cubic-form
invariant, it is easily seen that there exists no such closed Lie
algebra. This point holds true also for the algebraic generators
in various-power L-R invariants $I_{\rm l}^{n}$ ($n>3$). So, it is
concluded that for the driven oscillator, only the linear $I_{\rm
l}(t)$ and quadratic $I_{\rm q}(t)$ will form a complete set of
L-R invariants. Note that here $I_{\rm q}(t)$ should not be the
squared of $I_{\rm l}(t)$, {\it i.e.}, $I_{\rm q}(t)\neq cI_{\rm
l}^{2}(t)$, where $c$ is an arbitrary c-number. The existence of
$I_{\rm q}$ that cannot be written as the squared of any $I_{\rm
l}$ was demonstrated in Ref.\cite{Shen1}. It is emphasized here
that Bekkar {\it et al.}'s solution is the one constructed only in
terms of the eigenstates of the linear invariant $I_{\rm l}(t)$.
Even though for the linear invariant $I_{\rm l}(t)$ only, Bekkar
{\it et al.}'s result\cite{Bekkar} can truly be viewed as the
complete set of solutions (in a certain sub-Hilbert-space), it
still cannot be considered the general one of the Schr\"{o}dinger
equation, since the latter should contain those corresponding to
the quadratic invariant $I_{\rm q}(t)$. In brief, Bekkar {\it et
al.}'s solution and our solution, which will be found in what
follows, together constitute the complete set of solutions of the
Schr\"{o}dinger equation involving a time-dependent linear
potential.

In accordance with the L-R theory\cite{Lewis}, solving the
eigenstates of the quadratic invariant $I_{\rm q}(t)$ will enable
physicists to obtain the solutions of the time-dependent
Schr\"{o}dinger equation. But, unfortunately, it is not easy for
us to immediately solve the eigenvalue equation of the
time-dependent invariant $I_{\rm q}(t)$, for $I_{\rm q}(t)$
involves the time-dependent parameters. So, in the following we
will use the invariant-related unitary transformation
formulation\cite{Gao}, under which the {\it time-dependent}
invariant can be transformed into a {\it time-independent} one
$I_{V}$, and if the eigenstates of $I_{V}$ can be obtained
conveniently, the eigenstates of $I_{\rm q}(t)$ can then be easily
achieved.

For this aim, we will employ two time-dependent unitary
transformation operators
\begin{equation}
V_{1}(t)=\exp [\eta(t)q+\beta(t)p],   \quad
V_{2}(t)=\exp [\alpha(t)p^{2}+\rho(t)q^{2}]          \label{eq21}
\end{equation}
to get a {\it time-independent} $I_{V}$.  The time-dependent
parameters $\eta$, $\beta$, $\alpha$ and $\rho$ in (\ref{eq21})
are purely imaginary functions, which will be determined in what
follows\cite{Shen1}. Since the canonical variables (operators) $q$
and $p$ form a non-semisimple Lie algebra, here the first step is
to transform $I_{\rm q}(t)$ into $I_{1}(t)$, {\it i.e.},
$I_{1}(t)=V_{1}^{\dagger}(t)I_{\rm q}(t)V_{1}(t)$, which no longer
involves the canonical variables $q$ and $p$, and the retained Lie
algebraic generators in $I_{1}(t)$ are only $p^{2}$, $pq+qp$,
$q^{2}$. Here the time-independent parameters in $V_{1}(t)$ are
chosen\cite{Shen1}
\begin{equation}
\eta=\frac{EB'-FA'}{2i(E^{2}-DF)},    \quad
\beta=\frac{DB'-EA'}{2i(E^{2}-DF)}.
\end{equation}
Note that the three generators ($p^{2}$, $pq+qp$, $q^{2}$) in
$I_{1}(t)$ also form a Lie algebra\cite{Shen1}. The second step is
to obtain the {\it time-independent} $I_{V}$, which will be gained
via the calculation of $I_{V}=V_{2}^{\dagger}(t)I_{1}(t)V_{2}(t)$.
In this step, the obtained $I_{V}$ has no other generators (and
time-dependent c-numbers) than $p^{2}$ and $q^{2}$, namely,
$I_{V}$ may be written in the form $I_{V}=\frac{1}{2}\varsigma
\left (p^{2}+q^{2}\right)$ with $\varsigma$ being a certain
parameter independent of time\cite{Shen1}. For the detailed and
complicated derivation of the functions $\alpha$ and $\rho$,
readers may be referred to Ref.\cite{Shen1}. It is well known that
the eigenvalue equation of $I_{V}$ is of the form $I_{V}|n,
q\rangle=\left(n+\frac{1}{2}\right)\varsigma|n, q\rangle$, where
$|n, q\rangle$ stands for the familiar stationary
harmonic-oscillator wavefunction. Hence, the eigenstates of the
time-dependent L-R invariant $I_{\rm q}(t)$ can be achieved and
the final result is $V_{1}(t)V_{2}(t)|n, q\rangle$ with the
eigenvalue being $\left(n+\frac{1}{2}\right)\varsigma$.

According to the L-R invariant theory\cite{Lewis}, the particular
solution $\left| n; q, t\right\rangle _{\rm S}$ of the
time-dependent Schr\"{o}dinger equation is different from the
eigenfunction of the invariant $I_{\rm q}(t)$ only by a phase
factor $\exp \left[\frac{1}{i}\phi _{n}(t)\right]$, the
time-dependent phase of which is written as (in the unit
$\hbar=1$)
\begin{equation}
\phi _{n}(t)={\int_{0}^{t}}\langle n, q|
V^{\dagger}(t')\left[H(t')-i\frac{\partial}{\partial t'
}\right]V(t')|n, q\rangle {\rm d}t'
\end{equation}
with $V(t)=V_{1}(t)V_{2}(t)$. This phase $\phi _{n}(t)$ can be
calculated with the help of the Glauber formula and the
Baker-Campbell-Hausdorff formula\cite{Wei,EPJD}.

The particular solution $\left| n; q, t\right\rangle _{\rm S}$ of
the time-dependent Schr\"{o}dinger equation corresponding to the
invariant eigenvalue $\left(n+\frac{1}{2}\right)\varsigma$ is thus
of the form
\begin{equation}
\left| n; q, t\right\rangle _{\rm S}=\exp \left[\frac{1}{i}\phi
_{n}(t)\right]V_{1}(t)V_{2}(t)|n, q\rangle.
\end{equation}
Hence the general solution of the Schr\"{o}dinger equation
(corresponding to $I_{\rm q}(t)$) can be written in the form
\begin{equation}
|\Psi(q, t)\rangle_{\rm S}=\sum_{n}c_{n}\left| n; q,
t\right\rangle _{\rm S},
\end{equation}
where the time-independent c-number $c_{n}$'s are determined by
the initial conditions, {\it i.e.}, $c_{n}=\langle n,
q|V^{\dagger}(t=0)|\Psi(q, t=0)\rangle_{\rm S}$.

Thus we found the general solutions of the Schr\"{o}dinger
equation for the time-dependent linear potential, which
corresponds only to the quadratic-form invariant. As stated above,
Bekkar {\it et al.}'s solution is not the general one of the
Schr\"{o}dinger equation. Likewise, the solution obtained here
still does not form a complete set of solutions of this
time-dependent Schr\"{o}dinger equation, either. We conclude that
Bekkar {\it et al.}'s solution and our solution presented here
will together constitute such a complete set of solutions of the
Schr\"{o}dinger equation.

In addition, Guedes recently stated in his reply to Bekkar {\it et
al.} that in order to obtain the general solutions of the
time-dependent Schr\"{o}dinger equation one must follow the L-R
invariant theory {\it step by step}\cite{Guedes2}. I think that
``step by step'' may not be the essence of getting the general
solutions of Schr\"{o}dinger equation. Instead, the key point for
the present subject is that one should first find the complete set
of all L-R invariants of the time-dependent quantum systems under
consideration.


\begin{references}
\bibitem{Lewis}  H.R. Lewis, Jr. and W.B. Riesenfeld, J. Math. Phys.(N.Y) {\bf 10},
1458 (1969).
\bibitem{Guedes}  I. Guedes, Phys. Rev. A {\bf 63}, 034102 (2001).
\bibitem{Bekkar}  H. Bekkar, F. Benamira, and M. Maamache, Phys. Rev. A {\bf 68}, 016101 (2003)
\bibitem{Gao}  X.C. Gao, J.B. Xu, and T.Z. Qian, Phys. Rev. A {\bf 44}, 7016 (1991).
\bibitem{Shen1}     J.Q. Shen, arXiv: quant-ph/0310179 (2003).
\bibitem{Wei}     J. Wei and E. Norman, J. Math. Phys.(N.Y) {\bf 4}, 575
(1963).
\bibitem{EPJD} J.Q. Shen, H.Y. Zhu, and P. Chen, Euro. Phys. J. D
{\bf 23}, 305 (2003).
 \bibitem{Guedes2}  I. Guedes, Phys. Rev. A {\bf 68}, 016102
 (2003).
\end{references}
\end{document}